\documentclass[aip,jap,reprint,floatfix]{revtex4-1}

\usepackage{amsmath}
\usepackage{amssymb}
\usepackage{amsfonts}
\usepackage{graphicx}
\usepackage{epstopdf}

\newcommand{\uvz}{\mathbf{u}_z}

\newcommand{\uvx}{\mathbf{u}_x}
\newcommand{\uvr}{\mathbf{u}_r}
\newcommand{\uvrh}{\mathbf{u}_{\rho}}
\newcommand{\uvt}{\mathbf{u}_{\theta}}
\newcommand{\uvp}{\mathbf{u}_{\varphi}}

\newcommand{\ei}{\varepsilon_\text{i}}
\newcommand{\eeff}{\varepsilon_\text{eff}}

\newcommand{\erad}{\varepsilon_\text{rad}}
\newcommand{\etan}{\varepsilon_\text{tan}}
\newcommand{\er}{\varepsilon_r}
\newcommand{\et}{\varepsilon_\text{t}}
\newcommand{\erho}{\varepsilon_\rho}
\newcommand{\ephi}{\varepsilon_\varphi}
\newcommand{\epz}{\varepsilon_0}
\newcommand{\urad}{\mathbf{u}_\text{rad}}

\newcommand{\epsh}{\varepsilon_\text{h}}
\newcommand{\effpa}{\varepsilon_\text{eff,P}}
\newcommand{\effra}{\varepsilon_\text{eff,S}}
\newcommand{\alpa}{\alpha_\text{P}}
\newcommand{\alra}{\alpha_\text{S}}
\newcommand{\g}{\left(\frac{b}{a}\right)}

\newcommand{\vek}[1]{\mathbf{#1}}
\newcommand{\dyad}[1]{\overline{\overline{#1}}}

\begin{document}


\title{Cloaking and Magnifying Using Radial Anisotropy} 



\author{Henrik Kettunen}
\email[]{henrik.kettunen@helsinki.fi}
\affiliation{Department of Mathematics and Statistics, University of Helsinki, P.O.~Box 68, FI-00014, University of Helsinki, Finland}

\author{Henrik Wall\'en}
\affiliation{Department of Radio Science and Engineering, Aalto University School of Electrical Engineering, P.O.~Box 13000, FI-00076, AALTO, Finland}

\author{Ari Sihvola}
\affiliation{Department of Radio Science and Engineering, Aalto University School of Electrical Engineering, P.O.~Box 13000, FI-00076, AALTO, Finland}


\date{\today}

\begin{abstract}
This paper studies the electrostatic responses of a polarly radially anisotropic cylinder and a spherically radially anisotropic sphere. For both geometries, the permittivity components differ from each other in the radial and tangential directions. We show that choosing the ratio between these components in a certain way, these rather simple structures can be used in cloaking dielectric inclusions with arbitrary permittivity and shape in the quasi-static limit. For an ideal cloak, the contrast between the permittivity components has to tend to infinity. However, only positive permittivity values are required  and a notable cloaking effect can already be observed with relatively moderate permittivity contrasts. Furthermore, we show that the polarly anisotropic cylindrical shell has a complementary capability of magnifying the response of an inner cylinder. 
\end{abstract}


\maketitle 

\section{Introduction}
During the recent years, the concept of an electromagnetic invisibility cloak has been actively studied by mathematicians, physicists, and engineers alike. This has largely been due to the emergence of metamaterials research, having predicted that such a cloak could eventually be possible. The perhaps best known suggestions for designing an ideal cloak are based on transformation optics, \cite{Leonhardt06, Pendry06} where light is forced to go around the cloaked object without distortion. The corresponding coordinate transform had also been found slightly earlier related to electrical impedance tomography. \cite{Greenleaf03a, Greenleaf03b} However, the realization of this anisotropic and inhomogeneous cloak has proven very difficult.

Another famous cloaking approach is based on Mie scattering cancellation, \cite{Alu05} where a metamaterial coating is used to cancel out the dipolar field of a spherical object, so that in the long-wavelength limit, the coated object becomes completely invisible. The roots of this idea actually trace a couple of decades back in history. \cite{Kerker75, Chew76, Sihvola89, BohrenHuffman} Even though this method of \emph{plasmonic cloaking} \cite{Alu09} has shown to be rather robust against moderate perturbations of the inclusion geometry \cite{Alu07a} and it works also for several adjacent objects, \cite{Alu07b} the ideal coating must be designed separately for each inclusion with another size and different material parameters. The scattering cancellation approach has also been generalized for anisotropic spherical \cite{Gao08, Qiu09} and cylindrical \cite{Ni10} coatings and inclusions, increasing the degrees of freedom but also the complexity of the cloak design.

In this paper, we continue the study of anisotropic geometries and introduce an approximate and relatively simple quasi-static and non-magnetic cloaking approach based on \emph{radially anisotropic} (RA) permittivity. By radial anisotropy we mean that the considered geometries have clearly defined radial and tangential directions and their electric responses in these directions differ from each other. Radially anisotropic permittivity can be written in a dyadic form
\begin{equation}
\dyad{\varepsilon} = \epz\left[\erad\urad\urad + \etan\left(\dyad{I}-\urad\urad\right)\right],
\label{RA_eps}
\end{equation}
where $\epz$ is the permittivity of vacuum, $\erad$ and $\etan$ are the relative radial and tangential permittivities, respectively, $\dyad{I}$ is the unit dyadic, and $\urad$ is the unit vector in the radial direction. Note that Eq.~(\ref{RA_eps}) is independent of the coordinate system and the dimension of the geometry.

We consider two geometries that are special cases of radial anisotropy, namely a \emph{polarly radially anisotropic} (PRA) cylinder in 2D polar coordinates and a \emph{spherically radially anisotropic} (SRA) sphere in 3D spherical coordinates. These additional labels are introduced to retain the abbreviation RA general and coordinate-independent.

The electrostatic analysis of the PRA, or \emph{cylindrically anisotropic}, cylinder can be found in Ref.~\onlinecite{Yu06}. More often, even with a cylindrical geometry, the anisotropy has been considered with respect to Cartesian coordinates (see Ref.~\onlinecite{Ni10} and the references therein). In Ref.~\onlinecite{Jin10, Chen12}, scattering from (plasmonic) PRA cylinders is computed and, as already mentioned, Ref.~\onlinecite{Ni10} considers a PRA cylindrical shell for cloaking purposes.

Even more analysis on SRA spheres can be found. \cite{Schulgasser83, deMunck88, Helsing91, Sten95, MiltonKirja, Gao07} Such spheres have also been referred to using the term \emph{radially uniaxial}. \cite{Rimpilainen12, WallenEMTS13} Moreover, the general case, where all three components of $\dyad{\varepsilon}$ are allowed to be different has been investigated in Ref.~\onlinecite{Rimpilainen12}, where the sphere is called \emph{systropic}. Mie scattering from SRA spheres has been studied as well, \cite{Wong92, Qiu07, Qiu08, Qiu10} including the aforementioned cloaking studies. \cite{Gao08, Qiu09} A recent paper uses an SRA sphere as a model for a human head for brain imaging purposes. \cite{Petrov12} For an extensive list of occurrences and applications of radial anisotropy, see Ref.~\onlinecite{Gao08}. 

Herein, the analysis is based on quasi-electrostatics and finding the potential function $\phi(\vek{r})$ as a solution of the generalized Laplace equation 
\begin{equation}
\nabla \cdot \left(\dyad{\varepsilon}\cdot \nabla\phi\right) = 0.
\label{Laplace}
\end{equation} 

For both PRA cylinder and SRA sphere, we show that by choosing the contrast between $\erad$ and $\etan$ in a certain way, the structure can be used to cloak an inner inclusion in the quasi-static limit. Let us call this configuration an RA cloak. By letting the permittivity contrast tend to infinity, the RA cloak becomes ideal. Even though the approach is similar to plasmonic cloaking such that the cloaked inclusion is coated by another material layer, the principle of cloaking is different. The RA cloak does not give rise to a response opposite to the one of the inclusion to cancel it out, but it creates a zero electric field within itself allowing the hidden inclusion not to polarize at all. Therefore, the design of the RA cloak is fully independent of the shape and material of the cloaked inclusion. Moreover, the RA cloak is completely based on positive permittivities. In this sense, the RA cloak also resembles the cloaks achieved by transformation optics. The RA cloak can thus be seen as a simplification of the non-magnetic cloak suggested for optical frequencies. \cite{Cai07} 

The PRA cylinder can also perform a complementary operation. By inverting the permittivity ratio used for cloaking, it becomes a magnifying glass that transfers the response of the inner inclusion onto the boundary of the PRA shell making the inclusion effectively larger. This could prove an interesting discovery considering many sensing applications. Instead, the 3D SRA sphere does not share this characteristic.

In the following, we solve the polarizabilities and effective permittivities for structures where a dielectric inclusion, a cylinder or a sphere, is coated by cylindrical PRA or spherical SRA shell, respectively. It is assumed that all permittivity components are positive and the structures are surrounded by vacuum permittivity $\epz$. We further show how by tailoring the anisotropy ratio of the shell, the effective permittivity of the layered structure can be adjusted to the one of the surrounding space making the structure invisible, or in the 2D cylindrical case, alternatively to the one of the coated inclusion. Moreover, we provide a computational example verifying that the cloaking effect is independent of the inclusion shape.

\section{Polar Radial Anisotropy}
\begin{figure}[ht]
\includegraphics{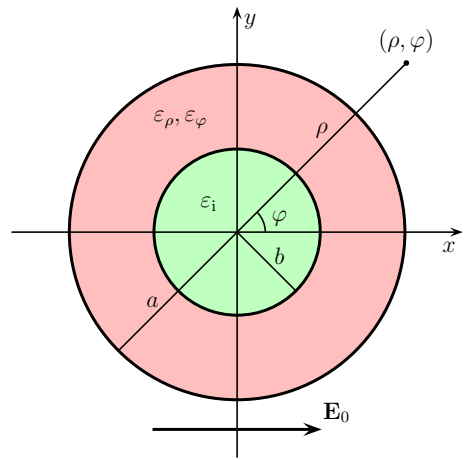}%
\caption{Cross-cut of an infinitely long circular cylindrical structure where an inner dielectric cylinder with radius $b$ and permittivity $\ei$ is covered by a polarly anisotropic layer with outer radius $a$ and permittivity components $\erho$ and $\ephi$. All permittivities are positive valued and given relative to the permittivity of vacuum $\epz$. The structure is excited by a uniform $x$-polarized static electric field $\vek{E}_0$. \label{PA_geom}} 
\end{figure}

\subsection{Polarizability and effective permittivity of a PRA-coated cylinder}\label{PA_A}
Let us first consider a cylindrical geometry with radial anisotropy restricted into two dimensions. In other words, we are only interested in the transverse response of an (infinitely) long straight cylinder, as the desired cloaking effect can only be seen using a transverse electric excitation. The cylinder may have an arbitrary axial permittivity component $\varepsilon_z$, but since the axial electric field is not excited, this component is omitted in our analysis. That is, we consider Eq.~(\ref{RA_eps}) the permittivity of a 2D disk with only one tangential, azimuthal, component  $\ephi$.

More precisely, let us study the layered structure presented in Fig.~\ref{PA_geom}, where a circular dielectric cylinder with radius $b$ and relative permittivity $\ei$ is coated by a PRA layer with outer radius $a$ and permittivity that is given in 2D polar coordinates as
\begin{equation}
\dyad{\varepsilon}_\text{P} = \epz\left(\erho\uvrh\uvrh + \ephi\uvp\uvp\right).
\label{PA_eps}
\end{equation}
Assume that this azimuthally symmetric structure is excited by an external uniform $x$-polarized electric field 
\begin{equation}
\vek{E}_0 = \uvx E_0 = \uvx\frac{U_0}{a}, 
\end{equation}
where $U_0$ is the potential difference across the cylinder radius $a$. The potential $\phi(\rho,\varphi)$ that satisfies the generalized Laplace equation (\ref{Laplace}) is of the form \cite{Yu06} 
\begin{subequations}
\begin{align}
\phi_\text{out} &= A\left(\frac{\rho}{a}\right)^{-1}\cos\varphi 
- U_0\left(\frac{\rho}{a}\right)\cos\varphi, \quad \rho \geq a \label{PA_out_pot}\\[1ex]
\phi_\text{P} &= B\left(\frac{\rho}{a}\right)^\gamma\cos\varphi
+C\left(\frac{\rho}{a}\right)^{-\gamma}\cos\varphi, \quad b \leq \rho \leq a \label{PA_pot}\\[1ex]
\phi_\text{in} &= D\left(\frac{\rho}{a}\right)\cos\varphi, \quad \rho \leq b, \label{PA_in_pot}
\end{align}
\end{subequations}
where 
\begin{equation}
\gamma = \sqrt{\frac{\ephi}{\erho}}.
\label{gamma}
\end{equation}

The coefficients $A$, $B$, $C$, and $D$ can be solved by applying the interface conditions on $\rho=a$ and $\rho=b$, and we can further solve the polarizability of the cylinder as (see Appendix \ref{AppA})
\begin{equation}
\alpa =2\frac{(\erho\gamma+1)(\erho\gamma-\ei)\g^{2\gamma}-(\erho\gamma-1)(\erho\gamma+\ei)}
{(\erho\gamma-1)(\erho\gamma-\ei)\g^{2\gamma}-(\erho\gamma+1)(\erho\gamma+\ei)}.
\label{alpa}
\end{equation}
The polarizability $\alpa$ is a dimensionless number describing the 2D transverse response of the structure. It is normalized by the vacuum permittivity $\epz$ and the cross-sectional area of the cylinder.

The corresponding polarizability of a homogeneous dielectric cylinder with relative permittivity $\epsh$ is
\begin{equation}
\alpha_\text{h} = 2\frac{\epsh-1}{\epsh+1}.
\end{equation}
This allows us to find an effective permittivity $\effpa$ for the coated cylinder such as
\begin{equation}
\alpa = 2\frac{\effpa-1}{\effpa+1},
\end{equation}
where
\begin{equation}
\effpa = \erho\gamma\frac{(\erho\gamma+\ei)-(\erho\gamma-\ei)\g^{2\gamma}}
{(\erho\gamma+\ei)+(\erho\gamma-\ei)\g^{2\gamma}}
\label{PA_eff_yleinen}
\end{equation}
This principle of \emph{internal homogenization} is discussed and applied in case of 3D coated spheres in Refs.~\onlinecite{Zeng10, Chettiar12, SihvolaEMTS13}. We note that due to the azimuthal symmetry, the internally anisotropic structure looks isotropic from the outside. 

\subsection{PRA cylinder as a cloak}\label{PA_B}
If we demand $\effpa=1$ in Eq.~(\ref{PA_eff_yleinen}), we obtain a design rule for an ideal PRA cloak, which is already given in Ref.~\onlinecite{Ni10}. We, however, consider a simpler approximative approach.

If the inner cylinder vanishes, we are left with an intact PRA cylinder, whose effective permittivity is obtained from Eq.~(\ref{PA_eff_yleinen}) by letting $b\to 0$ as
\begin{equation}
\effpa \to \erho\gamma = \sqrt{\erho\ephi}.
\label{sqrt}
\end{equation}
From Eq.~(\ref{sqrt}), we immediately see, like in Ref.~\onlinecite{Chen12}, that an intact PRA cylinder becomes invisible if
\begin{equation}
\erho\ephi=1.
\label{invisible_PA}
\end{equation}

Looking back at Eq.~(\ref{PA_eff_yleinen}), we notice that the inner cylinder can also be shrunk effectively, as the dependence of the radius $b$ is of the form $b^{2\gamma}$ and the exponent $\gamma$  in Eq.~(\ref{gamma}) is determined by the anisotropy ratio of the PRA shell. Let us choose the permittivities according to the invisibility condition of Eq.~(\ref{invisible_PA}) such as
\begin{equation}
\ephi = \kappa, \quad \erho = \frac{1}{\kappa},
\label{ephikappa}
\end{equation} 
where $\kappa$ is an arbitrary positive real number. The effective permittivity in Eq.~(\ref{PA_eff_yleinen}) becomes
\begin{equation}
\effpa = \frac{(\ei+1)+(\ei-1)\g^{2\kappa}}
{(\ei+1)-(\ei-1)\g^{2\kappa}}.
\label{PA_cloak_magn}
\end{equation}
Since $0 < b/a < 1$, the limit when the anisotropy ratio $\ephi/\erho = \kappa^2$ becomes infinite by $\kappa\to \infty$, gives
\begin{equation}
\effpa \to 1,
\end{equation}
and the structure becomes fully invisible. In other words, the PRA shell is capable of cloaking the inner cylinder regardless of the permittivity $\ei$ and the radius $b$.

It may not be possible to achieve the ideal cloaking condition in practice. However, with a finite $\kappa$, the PRA shell works as an approximate cloak. Let us study the 'worst case' considering the material of the inner cylinder, that is, cloaking a perfectly electrically conducting (PEC) cylinder. For a PEC inclusion with $\ei \to \infty$, the effective permittivity in Eq.~(\ref{PA_cloak_magn}) simplifies to
\begin{equation}
\effpa \to \frac{1+\g^{2\kappa}}{1-\g^{2\kappa}},
\end{equation}
which can further be written as
\begin{equation}
\effpa = 1 + 2\frac{\g^{2\kappa}}{1-\g^{2\kappa}} = 1 + \Delta\eeff.
\label{delta_PA}
\end{equation}
If we allow the effective permittivity of the nearly invisible structure to deviate at most the amount $\Delta\eeff$ from unity, we can solve the required $\kappa$ as
\begin{equation}
\kappa \geq \frac{\ln\left(\frac{\Delta\eeff}{\Delta\eeff+2}\right)}{2\ln\g}.
\label{PA_kappa}
\end{equation}

\begin{figure}[ht]
\includegraphics{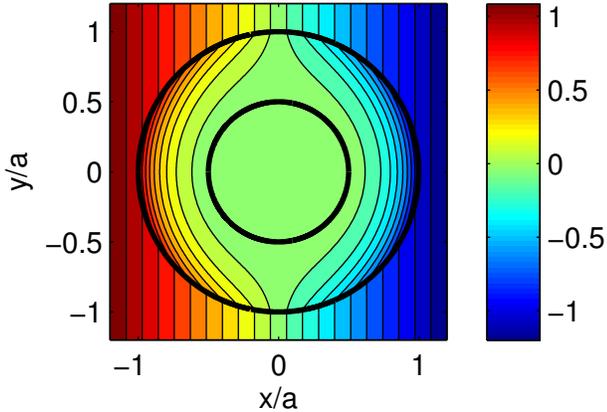}%
\caption{Potential distribution of a structure where a PRA annulus with azimuthal permittivity $\ephi = \kappa \approx 5.483$, radial permittivity $\erho = \kappa^{-1} \approx 0.182$ and outer radius $a$ cloaks a PEC cylinder with radius $b = a/2$ from an external $x$-polarized electric field. \label{PA_cloak}}
\end{figure}

For instance, if we have $b = a/2$ and we require $\Delta\eeff \leq 1\times 10^{-3}$, we need to choose the permittivity components as $\ephi = \kappa \approx 5.483$ and $\erho = \kappa^{-1} \approx 0.182$. This choice with relatively moderate anisotropy ratio $\ephi/\erho = \kappa^2 \approx 30$, is still enough to cloak the inner PEC cylinder almost completely. Figure \ref{PA_cloak} presents the potential distribution of the aforementioned situation in the $xy$ plane. The PRA annulus makes the potential levels bend such that a nearly constant zero potential and a zero field is formed around the origin, which is enough to hide the PEC cylinder. The polarizability of the structure becomes $\alpa \approx 1\times 10^{-3}$. 

In a more realistic quasi-static case, we also need to consider the effect of material losses. With lossy complex permittivities based on time convention $e^{j\omega t}$, the ideal invisibility condition (\ref{invisible_PA}) cannot be achieved without using an active material to compensate the negative imaginary part. With increasing losses the cloaking effect naturally begins to deteriorate. If we allow the permittivity deviation to become double, that is $|\Delta\eeff|\approx 2\times10^{-3}$, due to added losses in the previous example, where $\kappa \approx 5.483$, the azimuthal component can have an imaginary part as large as $\ephi \approx \kappa - j1.91\times10^{-2}$, when $\erho$ is assumed real. Unfortunately, the allowed losses in the radial component are much smaller as its real part is close to zero. To stay within the limits of the allowed deviation given above, the imaginary part of $\erho$ must remain smaller than $\erho = \kappa^{-1} -j6.26\times10^{-4}$, when $\ephi$ is assumed real.

If the cloak is designed to sufficiently cloak a PEC cylinder with radius $b$, it will cloak an inclusion, or a collection of inclusions, of any shape made of any dielectric material with $\varepsilon > 0$, as long as they altogether fit inside a cylindrical area with radius $b$. We demonstrate this by a computational example using \textsc{COMSOL Multiphysics} 4.3a, which is based on the finite element method (FEM).

\begin{figure}[ht]
\includegraphics{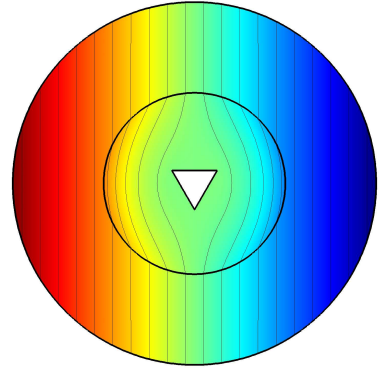}%
\caption{FEM simulation of the potential distribution in a case where a triangular PEC cylinder with side length $a/2$ is cloaked from an external $x$-polarized electric field using a cylindrical PRA coating with outer radius $a$ and permittivity components $\ephi = 2.5$ and $\erho = \ephi^{-1} = 0.4$ The outermost annulus consists of free space and the computational domain is terminated by a cylindrical boundary.   \label{Comsol_kolmio}}
\end{figure}

Let us consider cloaking a grounded triangular PEC cylinder whose side length is $a/2$, where $a$ is the outer radius of the applied cylindrical PRA coating. It turns out that the anisotropy ratio $\ephi/\erho = \kappa^2 = 6.25$ is already enough to cloak this inclusion sufficiently well. Figure \ref{Comsol_kolmio} presents the potential distribution where the inclusion is hidden from a $x$-polarized electric field.  For the normalized polarizability of the cloaked structure we obtain only $\alpa \approx 1.8 \times 10^{-3}$ whereas for the bare triangular inclusion we would have \cite{Thorpe92} $\alpha\approx 2.5811$. 

\subsection{PRA cylinder as a magnifying glass}
By saying that the PRA cylinder becomes a magnifying glass, we mean a situation where the properties of the inner dielectric cylinder are extended up to the outer surface of the PRA coating. This phenomenon can be seen in the case when there occurs a plasmonic resonance between the core and the coating, \cite{Nicorovici94, Nicorovici95, Yu06} that is, when $\erho\gamma = -\ei$. In the following, it is shown how this magnification is achieved with only positive permittivities regardless of the permittivity of the inclusion.

Let us further study the case where a dielectric inner cylinder is coated with a PRA shell with permittivity components chosen as in Eq.~(\ref{ephikappa}) and the effective permittivity of the structure is given by Eq.~(\ref{PA_cloak_magn}). If we consider the other limit when the permittivity ratio $\ephi/\erho$, instead of infinity, tends to zero as $\kappa \to 0$, we obtain
\begin{equation}
\effpa \to \ei.
\end{equation}
The structure then seems to be made completely out of the material of the inner cylinder. In other words, the PRA shell works as a magnifier that makes the inner cylinder radius effectively larger, in this ideal case up to $b \to a$.

With nonzero $\kappa$, Eq.~(\ref{PA_cloak_magn}) can be written as
\begin{equation}
\effpa = \ei - \frac{(\ei^2-1)-(\ei^2-1)\g^{2\kappa}}{(\ei+1)-(\ei-1)\g^{2\kappa}} = \ei - \Delta\eeff. 
\end{equation} 
If $\ei > 1$, $\Delta\eeff$ is positive indicating that with $\kappa > 0$, $\effpa$ underestimates $\ei$. Conversely, when $0 < \ei < 1$, $\Delta\eeff$ changes sign and with $\kappa \to 0$, $\effpa$ tends to $\ei$ from above. As it is convenient to consider the deviation $\Delta\eeff$ as a positive amount $|\Delta\eeff|$, we can write
\begin{equation}
\effpa = 
\begin{cases}
\ei - |\Delta\eeff|, & \ei > 1\\
\ei + |\Delta\eeff|, & 0 < \ei < 1. 
\end{cases}
\end{equation}
For the maximum allowed deviation $|\Delta\eeff|$, we must require
\begin{equation}
\kappa \leq \frac{\ln\tau}{2\ln\g},
\label{kappamagn}
\end{equation}
where
\begin{equation}
\tau = 
\begin{cases}
\frac{(\ei+1)(\ei-1-|\Delta\eeff|)}{(\ei-1)(\ei+1-|\Delta\eeff|)}, & \ei > 1\\[2ex]
\frac{(\ei+1)(\ei-1+|\Delta\eeff|)}{(\ei-1)(\ei+1+|\Delta\eeff|)}, & 0 < \ei < 1.
\end{cases}
\label{tau}
\end{equation}
Moreover, it is assumed that $|\Delta\eeff|$ is small compared to $\ei$, more precisely
\begin{equation}
|\Delta\eeff| < |\ei-1|.
\end{equation}

We note that successful magnifying requires much more extreme anisotropy ratios than it was required in the previous cloaking examples. If we consider magnifying a cylinder with radius $b = a/2$ and permittivity $\ei = 2$ with maximum deviation of one percent, $|\Delta\eeff| \leq 0.01$, according to Eqs.~(\ref{kappamagn}) and (\ref{tau}) we must have $\ephi = \kappa \approx 4.841\times 10^{-3}$ and $\erho = \kappa^{-1} \approx 206.6$, which gives the anisotropy ratio $\ephi/\erho \approx 2.34\times 10^{-5}$, or its inverse as large as $\erho/\ephi \approx 43000$. With the aforementioned parameter values, the polarizability of the structure becomes $\alpa \approx 0.662$. Since the normalized polarizability of a homogeneous cylinder with permittivity $\epsh = 2$ is $\alpha = 2/3$, the relative error in $\alpa$ is $0.7\%$.

\begin{figure}[ht]
\includegraphics{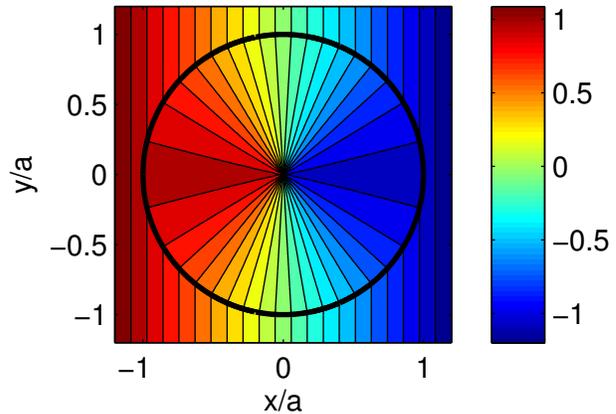}%
\caption{Potential distribution of an intact PRA cylinder with azimuthal permittivity $\ephi = \kappa \approx 4.841\times 10^{-3}$, radial permittivity $\erho = \kappa^{-1} \approx 206.6$ and outer radius $a$ in an external $x$-polarized electric field. The given anisotropy gives rise to a strong electric field in the origin. However, observed from the outside, the cylinder is invisible. \label{PA_magn_empty}}
\end{figure}

\begin{figure}[ht]
\includegraphics{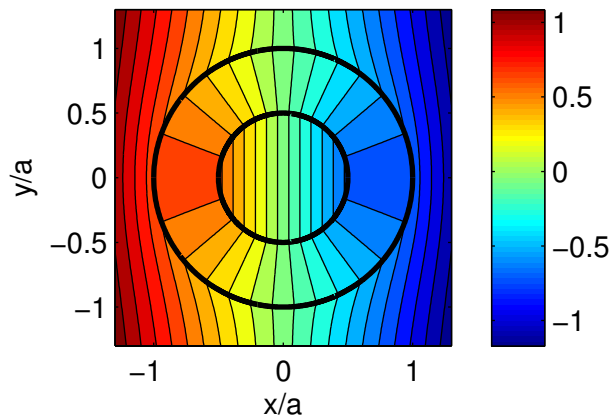}%
\caption{Potential distribution of a structure where a PRA annulus with azimuthal permittivity $\ephi = \kappa \approx 4.841\times 10^{-3}$, radial permittivity $\erho = \kappa^{-1} \approx 206.6$ and outer radius $a$ is used to magnify the response of a dielectric cylinder with radius $b = a/2$ and permittivity $\ei=2$ in an external $x$-polarized electric field. \label{PA_magn}}
\end{figure}

Figure \ref{PA_magn_empty} presents the potential distribution of an intact PRA cylinder in $x$-polarized electric field with permittivity components given above. Due to simultaneously large radial and small azimuthal permittivity the potential has a strong gradient at the origin. Although the cylinder is invisible observed from the outside, the structure is very sensitive to any perturbations near the origin. In Fig.~\ref{PA_magn}, a dielectric cylinder with $b=a/2$ and $\ei=2$ is inserted inside this PRA cylinder. We see that due to large $\erho$, at a given angle $\varphi$, the point at the surface on the inner cylinder, is (approximately) short-circuited to the outer surface of the PRA coating. On the other hand, $\ephi$ that is near zero preserves the potential distribution in the $\varphi$ direction. Altogether, observed from the outside, the structure resembles a homogeneous cylinder with radius $a$ and permittivity $\epsh \approx \ei = 2$.  

In the magnifying case, the effect of losses seems at first sight counterintuitive. Due to the required extreme permittivity contrast in the example above, the radial component is already so large that it begins to resemble a conducting medium from the viewpoint of the external electric field. Therefore, adding moderate losses to $\erho$ induces no significant effect. As our approximative magnifying glass with finite permittivity contrast underestimates the desired permittivity, we note that adding even larger losses to $\erho$ actually enhances the magnifying effect, the level of permittivity deviation finally saturating to $|\Delta\eeff|\approx 3.4\times10^{-3}$. On the other hand, introducing losses into the tangential component $\ephi$ with real $\erho$ makes the deviation grow becoming double the accepted level, $|\Delta\eeff| \approx 2\times10^{-2}$, with $\ephi \approx \kappa - j2.51\times10^{-2}$.    

\section{Spherical Radial Anisotropy}

\subsection{Polarizability and effective permittivity of an SRA-coated sphere}\label{RA_A}
Let us also consider the corresponding 3D spherical structure, where a dielectric sphere with permittivity $\ei$ and radius $b$ is coated with a spherical SRA shell with outer radius $a$. The permittivity of the SRA coating is given as
\begin{equation}
\dyad{\varepsilon}_\text{S} = \epz\left[\er\uvr\uvr + \et(\uvt\uvt+\uvp\uvp)\right].
\end{equation}
With an azimuthally symmetric $z$-polarized external electric field
\begin{equation}
\vek{E}_0 = \uvz E_0 = \uvz\frac{U_0}{a},
\end{equation}
the potential $\phi(r,\theta)$ that satisfies the Laplace equation (\ref{Laplace}) can be written as \cite{Schulgasser83, deMunck88, Helsing91, Sten95, MiltonKirja, Gao07}
\begin{subequations}
\begin{align}
\phi_\text{out} &= A\left(\frac{r}{a}\right)^{-2}\cos\theta 
- U_0\left(\frac{r}{a}\right)\cos\theta, \quad r \geq a \label{RA_out_pot}\\[1ex]
\phi_\text{S} &= B\left(\frac{r}{a}\right)^\nu\cos\theta 
+ C\left(\frac{r}{a}\right)^{-\nu-1}\cos\theta, \quad b \leq r \leq a \\[1ex]
\phi_\text{in} &= D\left(\frac{r}{a}\right)\cos\theta, \quad r \leq b \label{RA_in_pot}
\end{align}
\end{subequations}
where
\begin{equation}
\nu = \frac{1}{2}\left(-1+\sqrt{1+8\frac{\et}{\er}}\,\right).
\end{equation}
The effective permittivity of the structure becomes (see Appendix \ref{AppB})
\begin{equation}
\effra = 
\frac{\er\nu[\er(\nu+1)+\ei] - \er(\nu+1)(\er\nu-\ei)\g^\xi}
{\er(\nu+1)+\ei + (\er\nu-\ei)\g^\xi},
\label{RA_eff_yleinen}
\end{equation}
where 
\begin{equation}
\xi = \sqrt{1+8\frac{\et}{\er}} = 2\nu + 1.
\end{equation}

\subsection{SRA sphere as a cloak}
By letting the inner sphere vanish as $b \to 0$ in Eq.~(\ref{RA_eff_yleinen}), we obtain the effective permittivity of an intact SRA sphere as
\begin{equation}
\effra \to \er\nu.
\end{equation}
That is, an intact SRA sphere is invisible if 
\begin{equation}
\et = \kappa, \quad \er = \frac{1}{2\kappa-1},
\label{RA_invisible}
\end{equation}
where $\kappa \geq 1/2$ in order to keep both permittivity components positive . Choosing the parameters as above, the permittivity in Eq.~(\ref{RA_eff_yleinen}) becomes
\begin{equation}
\effra = \frac{[2\kappa+(2\kappa-1)\ei] + 2\kappa(\ei-1)\g^{4\kappa-1}}
{[2\kappa+(2\kappa-1)\ei]-(2\kappa-1)(\ei-1)\g^{4\kappa-1}}.
\label{RA_eff_cloak}
\end{equation}
Again, letting the ratio $\et/\er = \kappa(2\kappa-1)$ grow as $\kappa \to \infty$, yields
\begin{equation}
\effra \to 1,
\end{equation}
and similarly to the PRA cylinder, the SRA sphere works as a cloak. 

The magnification of the inner core can again be achieved by the plasmonic condition, \cite{Nicorovici95} $\er\nu = -\ei/2$. Instead, an SRA magnifying glass cannot be constructed using positive permittivity components. First of all, reaching the limit $\kappa \to 0$ would require $\er$ to be negative. Moreover, this limit does not even yield the magnifying operation, as then $b^{4\kappa-1} \to b^{-1}$ and the solution remains dependent on the radius $b$. The limit $b^{4\kappa-1} \to 0$ would be obtained when $\er = -2$ and $\et = 1/4$, but within the scope of this paper, we do not consider negative permittivity components, as they would require more intricate analysis. \cite{WallenEMTS13}

\subsection{Comparison between cylindrical and spherical structures}
\begin{figure}[ht]
\includegraphics{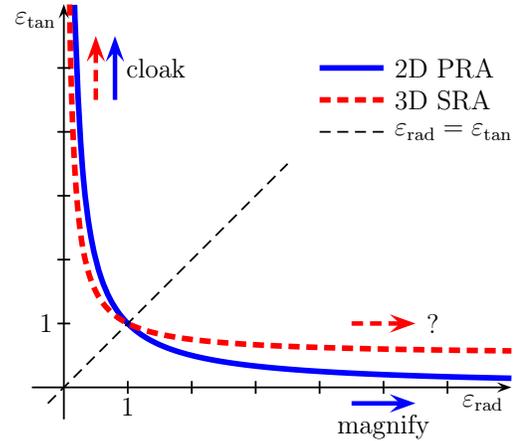}%
\caption{The anisotropy ratios that make the intact PRA cylinder and SRA sphere invisible, Eqs.~(\ref{ephikappa}) and (\ref{RA_invisible}), respectively, plotted on a linear scale. With $\etan/\erad \to \infty$ both structures can be used for cloaking, whereas only the PRA cylinder works as a magnifier when $\etan/\erad \to 0.$ \label{PA_RA_kappa}}
\end{figure}
\begin{figure}[ht]
\includegraphics{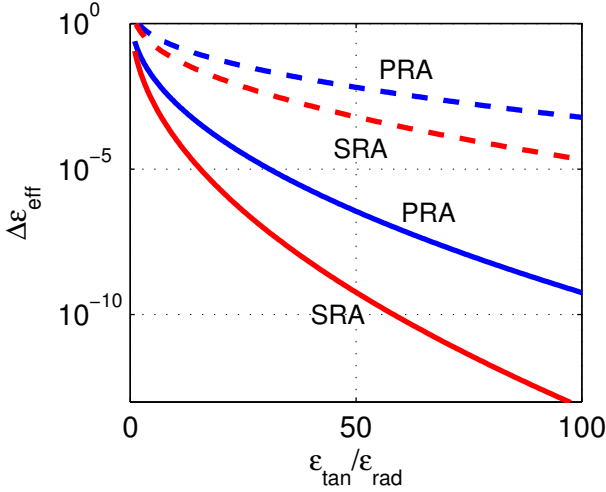}%
\caption{Comparison between the PRA cloak (blue lines) and the SRA cloak (red lines) when cloaking a PEC inclusion. The deviation $\Delta\eeff$ from ideal invisibility $\eeff = 1$, for PRA given by Eq.~(\ref{delta_PA}) and for SRA by Eq.~(\ref{delta_RA}), is plotted as a function of the anisotropy ratio $\etan/\erad$ for two radius ratios $b = a/3$ (solid lines) and $b = 2a/3$ (dashed lines). \label{PA_RA_g3g6}}
\end{figure}
\begin{figure}[ht]
\includegraphics{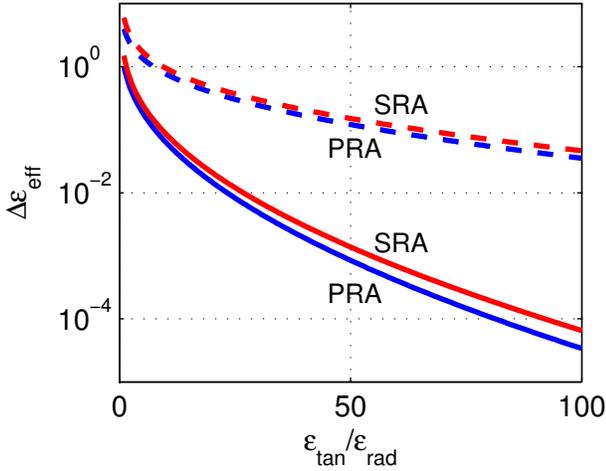}%
\caption{Comparison between PRA (blue lines) and SRA (red lines) cloaks as deviation $\Delta\eeff$ as in Fig.~(\ref{PA_RA_g3g6}). The curves are plotted for inclusion filling factors $f_\text{P} = f_\text{S} = f$ of $f=1/3$ (solid lines) and $f=2/3$ (dashed lines). \label{PA_RA_F3F6}}
\end{figure}

The comparison between the 2D PRA and 3D SRA structures reveals both similarities and differences. Figure \ref{PA_RA_kappa} presents the permittivity ratios between the tangential component $\etan$ and the radial component $\erad$ that satisfy the invisibility conditions of Eqs.~(\ref{ephikappa}) and (\ref{RA_invisible}), for an intact PRA cylinder and SRA sphere, respectively. Given as a function of $\etan = \kappa$, the anisotropy ratios $\etan/\erad$ become in the PRA case $\ephi/\erho = \kappa^2$ and in the SRA case $\et/\er = \kappa(2\kappa-1)$. With simultaneously increasing tangential and decreasing radial permittivity, both structures can be used as a cloak. On the other hand, when the ratio $\etan/\erad$ tends to zero, only the PRA cylinder has the capability to magnify the response of the inner inclusion. 

We can construct both approximate PRA and SRA cloaks with relatively moderate anisotropy ratios. Let us compare their performance in cloaking a PEC inclusion. The required parameter $\kappa$ for the PRA cloak to achieve deviation smaller than $\Delta\eeff$ from the invisibility $\eeff = 1$ is already given in Eq.~(\ref{PA_kappa}). For the SRA cloak, the effective permittivity in Eq.~(\ref{RA_eff_cloak}) with a PEC inclusion, $\ei \to \infty$, can be written as
\begin{equation}
\effra \to 1 + \Delta\eeff = 1 + \frac{(4\kappa-1)\g^{4\kappa-1}}{(2\kappa-1)\left[1-\g^{4\kappa-1}\right]}.
\label{delta_RA}
\end{equation}    
In this case, the parameter $\kappa$ does not have a simple explicit solution.

The deviation $\Delta\eeff$ is plotted as a function of the anisotropy ratio $\etan/\erad$ for the PRA and SRA cloaks with two radius ratios, $b/a = 1/3$ and $b/a = 2/3$, in Fig.~\ref{PA_RA_g3g6}. In this comparison, the SRA cloak seems to be notably more efficient by giving the same $\Delta\eeff$ with smaller anisotropy ratio than the PRA cloak. We must, however, note that with a given $b/a$, the fraction that a 3D inclusion occupies out of the total volume is smaller than the area fraction of the corresponding 2D inclusion. Let us therefore define a filling factor $f$, which for the PRA cloak is the square, $f_\text{P} = (b/a)^2$, and for the SRA cloak the cube, $f_\text{S} = (b/a)^3$, of the radius ratio. When the $\Delta\eeff$ curves are plotted demanding the same filling factors for both cloaks, as in Fig.~\ref{PA_RA_F3F6} with $f = 1/3$ and $f = 2/3$, the performance of the cloaks becomes much more equal, the PRA cloak being slightly better in this comparison.

According to computational comparison, with equal filling factors, the SRA cloak's sensitivity to losses appears to be on the same order of magnitude as the PRA cloak described in Section \ref{PA_B}. However, in the SRA case, the required permittivity contrast is achieved with smaller absolute values of $\erad$ and $\etan$ than in the PRA case making the SRA structure slightly more vulnerable to increasing imaginary parts. 

\section{Conclusion}
In this paper, we have studied the responses of radially anisotropic (RA) geometries in a uniform static field. We have shown that a 2D polarly radially anisotropic (PRA) cylinder and a 3D spherically radially anisotropic (SRA) sphere with moderately large anisotropy ratios $\etan/\erad$ can be used in cloaking dielectric inclusions of arbitrary shape. In addition, the PRA cylinder starts to behave as a magnifying glass for the inner inclusion when the anisotropy ratio tends to zero, whereas the spherical structure does not share this characteristic. 

The cloaking is based on a certain relation between the radial and tangential components, which makes the intact structure invisible. When the tangential component then tends to infinity, the radial one tends to zero. This gives rise to a strong gradient in the potential on the interface between the RA material and free space, whereas the interior of the RA cloak remains in the zero potential creating a region where any dielectric object can be inserted without being observed from the outside. That is, one cloak design works for all inclusions independent of their permittivity, shape or number. In other words, the ideal RA cloak is a special kind of Faraday cage that is also invisible.

Our analysis of the cloak is, however, based on electrostatics and its ideal performance can so far only be verified when cloaking non-magnetic objects from a static electric field. To achieve cloaking in the case of magnetic inclusions with magnetostatic excitation, the permeability $\dyad{\mu}$ of the cloak must be made radially anisotropic. A crucial future objective is to study the performance of the RA cloak in a dynamic case with propagating electromagnetic waves. We may, however, assume that similarly to Refs.~\onlinecite{Alu05, Gao08, Qiu09, Ni10}, the cloaking effect extends to the long-wavelength region, where the cloak is not extremely small compared with the wavelength. The suggested RA cloak may prove useful in optical frequencies where magnetism is not an issue. A phenomenon that clearly affects the performance of the cloak destructively is the first magnetic Mie resonance, which sets a certain limit for the maximum electrical size of the cloak. 

We have estimated the effect of material losses by a couple of computational examples. The full-wave frequency domain analysis is still needed to give more detailed insight on the RA cloak's sensitivity to material dispersion and losses and to further distinguish the effects of absorption and scattering. Since the operating principle of the cloak is not based on a tuned resonant behavior of the structure, very small losses do not immediately eradicate the cloaking effect. However, increasing imaginary parts of the permittivity components monotonically decrease the performance of the cloak. Especially the radial permittivity component, whose real part is required to be near zero, becomes very sensitive to material losses. 

There are at least two different ways to construct an object with radial anisotropy, let us call them the hedgehog and the onion approaches. The hedgehog structure is constructed using radially oriented wires or needles as it is suggested in Ref.~\onlinecite{Cai07}. The onion structure is based on multiple concentric layers. In the 2D case we may consider a cross-cut of a tree trunk with concentric annual rings. The simplest realization of an RA cloak is perhaps a cylindrical cloak, which consists of multiple isotropic layers. This kind of geometry is considered, for example, in Ref.~\onlinecite{Tricarico09}. There must be enough layers and they must be thin enough so that the material can be considered effectively homogeneous. The radial and tangential components of the RA cloak become tunable being functions of the permittivities and thicknesses of the layers. On the other hand, achieving a permittivity near zero calls for a some kind of metamaterial realization. 
We must also note that the cylindrical structure is designed only for the case where the polarization of the electric field is transverse to the cylinder axis.

The magnifying behavior of the PRA cylinder may also prove useful considering sensing applications. Successful magnification, however, requires an extreme contrast between the permittivity components and the cylindrical shape of the studied inclusion.

\begin{acknowledgments}
The work of H. Kettunen was supported by the Academy of Finland projects 260522 and CoE-250215.
\end{acknowledgments}

\appendix
\section{PRA-coated cylinder in a static electric field} \label{AppA}
The solution for the generalized Laplace equation (\ref{Laplace}) in a PRA medium with permittivity given by Eq.~(\ref{PA_eps}) can be found in Ref.~\onlinecite{Yu06}, where also the case of a layered cylinder is considered. The structure described in Section \ref{PA_A} with an $x$-polarized excitation thus yields the solutions (\ref{PA_out_pot})--(\ref{PA_in_pot}). The interface conditions become with $\rho=a$
\begin{align}
A - U_0 &= B + C \label{eka}\\
-A - U_0 &= \erho\gamma B - \erho\gamma C
\end{align}
and with $\rho=b$
\begin{align}
\g^\gamma B + \g^{-\gamma}C &= \g D\\
\erho\gamma\g^{\gamma-1}B - \erho\gamma\g^{-\gamma-1}C &= \ei D. \label{vika}
\end{align}
From Eqs.~(\ref{eka})--(\ref{vika}), we can solve
\begingroup
\allowdisplaybreaks
\begin{align}
A &= U_0\frac{(\erho\gamma+1)(\erho\gamma-\ei)\g^{2\gamma}-(\erho\gamma-1)(\erho\gamma+\ei)}
{(\erho\gamma-1)(\erho\gamma-\ei)\g^{2\gamma}-(\erho\gamma+1)(\erho\gamma+\ei)} \label{kerroin_A}\\[2ex]
B &= U_0\frac{2(\erho\gamma+\ei)}
{(\erho\gamma-1)(\erho\gamma-\ei)\g^{2\gamma}-(\erho\gamma+1)(\erho\gamma+\ei)}\\[2ex]
C &= U_0\frac{2(\erho\gamma-\ei)\g^{2\gamma}}
{(\erho\gamma-1)(\erho\gamma-\ei)\g^{2\gamma}-(\erho\gamma+1)(\erho\gamma+\ei)}\\[2ex]
D &= U_0\frac{4\erho\gamma\g^{\gamma-1}}
{(\erho\gamma-1)(\erho\gamma-\ei)\g^{2\gamma}-(\erho\gamma+1)(\erho\gamma+\ei)}.
\end{align}
\endgroup

Let us define the 2D normalized polarizability of the cylindrical structure $\alpa$ as a ratio between the induced dipole moment $p$ and the external electric field $\vek{E_0}$ such as
\begin{equation}
\vek{p} = \pi a^2\epz\alpa \vek{E}_0.
\end{equation} 
Since the potential given rise by a 2D $x$-polarized line dipole, $\vek{p} = \uvx p$, is of the form
\begin{equation}
\phi_\text{d} = \frac{p\cos\varphi}{2\pi\epz\rho},
\end{equation}
we obtain the polarizability from the coefficient $A$ given by Eq.~(\ref{kerroin_A}) as
\begin{equation}
\alpa = 2\frac{A}{U_0},
\end{equation}
which further yields Eq.~(\ref{alpa}).
\section{SRA-coated sphere in a static electric field} \label{AppB}
The solution for the Laplace equation in an SRA medium can be found in e.g.~Refs.~\onlinecite{Schulgasser83, deMunck88, Helsing91, Sten95, MiltonKirja, Gao07}. Ref.~\onlinecite{Gao07} also considers the layered case. Considering the coated spherical structure described in Section \ref{RA_A}, we obtain the solution (\ref{RA_out_pot})--(\ref{RA_in_pot}), whose coefficients need to be solved from the interface conditions. 

On the interface with $r=a$
\begin{align}
A - U_0 &= B + C \label{RA_eka}\\
-2A - U_0 &= \er\nu B - \er(\nu+1)C,
\end{align}
and on the interface with $r=b$
\begin{align}
\g^\nu B - \g^{\nu+1}C &= \g D\\
\er\nu \g^{\nu-1}B - \er(\nu+1)\g^{-\nu-2}C &= \ei D. \label{RA_vika}
\end{align}
From Eqs.~(\ref{RA_eka})--(\ref{RA_vika}) we can solve the coefficients 
\begin{widetext}
\begin{align}
A &= U_0
\frac{(\er\nu-\ei)[\er(\nu+1)+1]\g^{2\nu+1} - (\er\nu-1)[\er(\nu+1)+\ei]}
{(\er\nu-\ei)[\er(\nu+1)-2]\g^{2\nu+1} - (\er\nu+2)[\er(\nu+1)+\ei]} \label{RA_kerroin_A}\\[2ex]
B &= U_0
\frac{3[\er(\nu+1)+\ei]} 
{(\er\nu-\ei)[\er(\nu+1)-2]\g^{2\nu+1} - (\er\nu+2)[\er(\nu+1)+\ei]}\\[2ex]
C &= U_0
\frac{3(\er\nu-\ei)\g^{2\nu+1}}
{(\er\nu-\ei)[\er(\nu+1)-2]\g^{2\nu+1} - (\er\nu+2)[\er(\nu+1)+\ei]}\\[2ex]
D &= U_0
\frac{3\er(2\nu+1)\g^{\nu-1}}
{(\er\nu-\ei)[\er(\nu+1)-2]\g^{2\nu+1} - (\er\nu+2)[\er(\nu+1)+\ei]}. \label{RA_kerroin_D}
\end{align}
\end{widetext}
Let us now define the 3D normalized polarizability of the spherical structure, $\alra$, by the relation
\begin{equation}
\vek{p} = \frac{4}{3}\pi a^3 \epz \alra \vek{E}_0.
\end{equation}
The potential of a $z$-polarized dipole, $\vek{p} = \uvz p$, becomes
\begin{equation}
\phi_\text{d} = \frac{p\cos\theta}{4\pi\epz r^2},
\end{equation}
and the polarizability is obtained from the coefficient A given by Eq.~(\ref{RA_kerroin_A}) as
\begin{equation}
\alra = 3\frac{A}{U_0}.
\end{equation}
The effective permittivity, Eq.~(\ref{RA_eff_yleinen}), is derived with straightforward algebra using the equation of the polarizability of a homogeneous sphere such as
\begin{equation}
\alra = 3\frac{\effra-1}{\effra+2}.
\end{equation}

\end{document}